# Continuum approach based on radiation distribution function for radiative heat transfer in densely packed particulate system

Baokun Liu[a], Junming Zhao[a,c,*], Linhua Liu[b]

*[a] School of Energy Science and Engineering, Harbin Institute of Technology, Harbin 150001, China*
*[b] School of Energy and Power Engineering, Shandong University, Qingdao 266237, China*
*[c] Key Laboratory of Aerospace Thermophysics, Ministry of Industry and Information Technology of China*

## Abstract

The study of radiative heat transfer in particulate system is usually based on radiative transfer equation (RTE) with effective radiative properties. However, for non-random, densely and regularly packed particulate systems, the applicability of RTE is questionable due to dependent scattering and weak randomness of particle arrangement. In this paper, a new continuum approach that does not explicitly rely on the RTE is proposed for radiative heat transfer in the densely packed particulate system. The new approach is based on the generalization of the concept of radiation distribution factor (RD) at discrete scale (or particle scale) to radiation distribution function (RDF) at continuum scale. The derived governing equation is in integral form, with RDF as the continuum scale physical parameter that characterize the radiative transfer properties of the system. The characteristics of the RDF for different particulate system (randomly or regularly packed) are studied. The RDF in regularly distributed particulate system is shown to be anisotropic. The continuum approach is verified by using heat transfer simulation at the particle scale, and demonstrated to have excellent performance in predicting the temperature distribution in dense particulate system. This work provides an alternative continuum theory, which will help the analysis and understanding of radiative heat transfer in densely packed media.

---

* Corresponding author.
*Email address*: jmzhao@hit.edu.cn (Junming Zhao)

**Nomenclature**

| | |
|---|---|
| $A$ | surface area, m$^2$ |
| $a, b, c$ | coefficient |
| $C$ | heat capacity, J/K |
| $c_p$ | specific heat, J/(kg·K) |
| $f$ | particle filling ratio |
| $i, j$ | number index |
| $n, k$ | optical constant |
| $N$ | number of particles |
| $p', q'$ | parameters defined by Eq. (41) |
| $Q$ | power, W |
| $r$ | distance, m |
| $R$ | radius of particle, m |
| $RD$ | radiation distribution factor |
| $RDF$ | radiation distribution function |
| $T$ | temperature, K |
| $t$ | time, s |
| $V$ | computational domain, volume, m$^3$ |
| $\mathbf{x}, \mathbf{x}'$ | position |
| $x$ | size parameter, $x$ coordinate |
| $y$ | $y$ coordinate |
| $z$ | $z$ coordinate |

*Greek Symbols*

| | |
|---|---|
| $\alpha$ | absorptivity |
| $\beta, \gamma$ | fitting parameter defined in Eq. (45) |
| $\delta$ | particle spacing, m |
| $\varepsilon$ | wall emissivity |
| $\theta, \varphi$ | zenith and azimuth angle |
| $\lambda$ | wavelength, μm |
| $\xi$ | random number |
| $\sigma$ | Stefan-Boltzmann constant, $5.67 \times 10^{-8}$ W/(m$^2$·K$^4$) |
| $\rho$ | density, kg/m$^3$ |
| $\rho_N$ | number density of particles, 1/m$^3$ |
| $\rho_R$ | reflectivity |

*Subscripts*

| | |
|---|---|
| Bnd | boundary region |
| *cell* | cellular |
| *div* | volume element |
| *eff* | effective parameter |
| $E$ | self-emission |
| $En$ | enclosure wall |
| $i, j$ | serial number |
| *in* | incident |
| Int | interior region |
| $L$ | left |
| $P$ | particle |
| $R$ | right |
| $r$ | reflect |
| $S$ | Source heating |
| $\parallel$ | parallel polarized |
| $\perp$ | perpendicular polarized |

*Superscripts*

| | |
|---|---|
| PP | from particle to particle |
| PdS | from particle to differential surface element |
| SP | from surface to particle |
| * | dimensionless parameter |

# 1  Introduction

In nature and industrial processes, many substances, products (including intermediate products) are in particulate state. Radiative heat transfer in particulate system has been an extensive researched issue in many science and technology fields, such as atmospheric science [1], solar energy [2, 3], remote sensing [4], nuclear energy [5], aerospace [6], packed beds and fluidized beds [7].

The most common approach to solve the radiative transfer problem in dispersive media is to solve the radiation transfer equation (RTE). Many methods of the RTE solution have been developed [8-10] and a short review on the numerical solution of RTE can be found in Ref. [11]. However, there are several assumptions in the derivation of RTE from first principles [12, 13], such as the particles are located in the far-field zones of each other, the scatterers are considered as point scatterers and the position of particles are completely uncorrelated [14].

For dilute dispersive media, those assumptions are usually fulfilled. The radiation interaction with particles can be viewed as 'point scattering' [15], i.e. the particles act like point scatterers with negligible volume and they act independently in light scattering. In this situation, the interaction of radiation with an isolated particle is called 'independent scattering' [16-18]. However, when the filling ratio of scatterers is so high that particles are closely spaced, particles do not scatter independently and there will exist the phenomena known as "dependent scattering" [15-18]. It includes the 'interference effects', i.e. the interaction of individual particles with radiation will be influenced by other particles in proximity; 'multiple scattering effects', i.e. the scattered radiation by one particle is incident on another particle to be scattered again, and the 'volume scattering' effect, which is proposed by Brewster [19], i.e., the particles are no longer regarded as point scatterers. Besides, when the particles are tightly packed, the positions of particles will be correlated with each other and tends to be regular. Therefore, in such a non-random medium, radiation transfer becomes more complicated [20] and the RTE may fail to correctly predict the radiative heat transfer [13, 21-23].

For radiation in densely packed systems or porous materials, the application of the RTE is questionable due to the largely violation of the theoretical assumptions. Even though, the RTE with "effective radiative properties" is usually applied to analyze radiative heat transfer in such system. In this case, the determination of "effective radiative properties" is the main task [15]. For a densely packed system consisting of optically large particles, the geometric optics approximation (GOA) by neglecting wave effects is valid [24-26]. For packed beds of opaque particles, the radiative properties considering the dependent scattering effect can be considered by scaling the independent radiative properties [19, 27, 28]. A Monte Carlo technique were developed to directly determine the effective radiative properties of porous media [29, 30]. Recently, Randrianalisoa and Baillis [15] developed an approach for calculating effective radiative properties based on the mean-free-paths theory. They also presented an analytical formulation of radiative properties to reduce the computation time of the Monte Carlo algorithm [26]. Besides the direct application of the classic RTE to porous media, several schemes on the modification of the classic RTE that are considered more suitable to porous media were also proposed. Gusarov [31] proposed a vector RTE to study radiation transfer in heterogeneous media.

In this scheme, the radiation in each phase is described by a separate RTE and those RTEs are coupled with each other by exchanging terms. Some drawbacks of this approach were pointed out in Ref. [32]. Lipinski et al. [33, 34] developed a multiphase radiative transfer theory based on the spatial averaging theorem for multiphase heterogeneous media. Randrianalisoa et al. [35] compared the classic RTE and multiphase RTE approaches in the random fiber media and obtained the scope of application of these two approaches.

Radiative heat transfer in non-random media is more complicate, since radiative transfer may be anisotropic and the Beer's law may not be valid anymore. Bellet et al. [36] studied the anisotropic and scattering dependent radiative conductivity tensors in porous media with application to radiative heat transfer in rod bundles. Taine et al. [37] developed a new theory, i.e., the generalized RTE, for radiative heat transfer in non-Beerian media. This approach was further applied to strongly non-homogeneous porous media with application to a slab of packed particles [38] and non-Beerian effective phases of macroporous media [39]. Dayal et al. [40] used view factors to analyze radiative heat transfer in granular medium at particle scale. Since practical system contains a large number of particles, the computation time and memory consumption for radiative heat transfer analysis directly at the particle scale is huge. Most recently, Wu et al. [23, 41] proposed a novel continuum approach, i.e., the approximation function model for solving radiative heat transfer in densely packed bed, in which radiation exchange is characterized by view factors. Until now, there are only very few works reported on continuum approach that doesn't explicitly rely on the RTE for radiative heat transfer in the densely packed particulate system, and the understanding of radiation transfer in non-random system is also very limited.

In this work, a continuum approach that does not explicitly rely on RTE is proposed to solve the radiative heat transfer in non-random particle system. The new approach is based on the generalization of the concept of RD at particle scale to RDF at continuum scale. The obtained governing equation is in integral form, with RDF as continuum scale physical parameter. The characteristics of the RDF for different particulate system (randomly or regularly packed) are studied. The performance of the continuum approach to simulate the temperature distributions in a non-random arrangement particulate system is demonstrated by comparison with heat transfer simulation at the particle scale.

## 2 Theoretical development

### 2.1 Particle scale energy balance in densely particulate system

In order to investigate radiation heat transfer in the particulate system, a system that contains a large number of particles is considered, as depicted in Fig. 1. For simplicity, in the following analysis, it is assumed that all the particles are optically large (with size parameter $x \doteq \pi d/\lambda$ is much larger than 1) and the geometric optics approximations is valid.

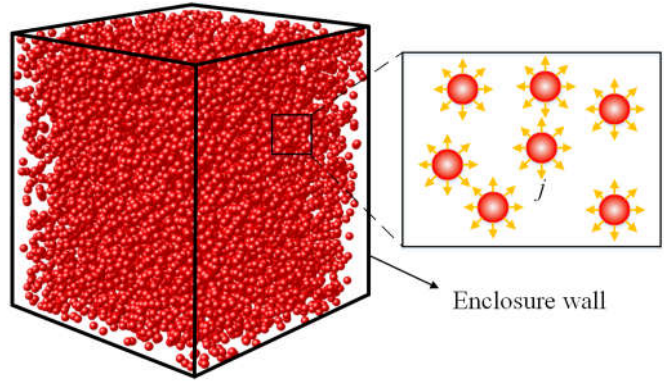

**Fig 1.**    Schematic diagram of the radiation heat transfer in optically large particle system.

Considering radiative heat transfer among particles and enclosure walls, and other heat source like laser heating or convection heat transfer with environment, the transient energy balance equation for a given particle (with number $j$) in the system can be formally written as follows.

$$C_j \frac{\partial T_j}{\partial t} = Q_{P,j} + Q_{En,j} - Q_{E,j} + Q_{S,j} \tag{1}$$

where $C_j = (\rho c_p V_P)_j$ is the heat capacity, $T_j$ is the temperature, $Q_{P,j}$ and $Q_{En,j}$ denote the radiation flux contributed by neighboring particles and enclosure walls (W), respectively, $Q_{E,j} = A_j \varepsilon_j \sigma T_j^4$ is the self-emission power (W), in which $\varepsilon_j$ and $A_j$ denote the emissivity and surface area of the particle, $Q_{S,j}$ denotes source heating power (W).

Detailed radiative heat transfer analysis must be employed in order to get explicitly form of $Q_{P,j}$ and $Q_{En,j}$. Here, the mathematic form of $Q_{P,j}$ and $Q_{En,j}$ will be derived based on the concept of radiation distribution factor (RD) [42]. The RD between two interaction particles, $RD_{i,j}$, is defined as the ratio of the absorbed power by the target-particle $j$ to the total emitted power from the source-particle $i$ [42], which can be calculated from

$$RD_{i,j} = \frac{Q_{i,j}}{\varepsilon_i A_i \sigma T_i^4} \qquad (2)$$

where $Q_{i,j}$ denotes the radiative power absorbed by particle $j$ that is emitted by particle $i$.

As compared to the concept of view factor (or configuration factor), the value of RD depends not only on geometry of particle arrangement, but also on the radiative properties of interaction elements, i.e., particles and enclosure [43], and radiative interactions, such as multiple scattering or reflections. Hence RD totally characterizes radiative heat transfer. Note that the concept of RD is general, which can be applied to particles of any shape and with any kind of surface properties, such as diffuse or Fresnel reflection. In case the self-emissive power of a particle is well defined, the RD can be defined.

With the help of RD, the $Q_{P,j}$ and $Q_{En,j}$ can be calculated from

$$Q_{P,j} = \sum_{i=1}^{N_P} A_i \varepsilon_i \sigma T_i^4 RD_{i,j}^{\mathrm{PP}} \qquad (3)$$

$$Q_{En,j} = \sum_{i=1}^{N_{En}} A_i \varepsilon_i \sigma T_i^4 RD_{i,j}^{\mathrm{SP}} \qquad (4)$$

where $RD_{i,j}^{\mathrm{PP}}$ denotes the particle-particle (PP) RD from particle $i$ to particle $j$, $RD_{i,j}^{\mathrm{SP}}$ denotes the surface-particle (SP) RD from surface element $i$ of enclosure wall to particle $j$, $N_P$ denotes the total number of particles, $N_{En}$ denotes of the total number of surface elements of the enclosure wall.

The RD has reciprocity and conservation properties [42], which are the results of the reversibility of optical path and conservation of emitted power, and are similar to the properties of the view factor, namely,

$$\varepsilon_i A_i RD_{i,j} = \varepsilon_j A_j RD_{j,i} \qquad (5)$$

$$\sum_{j}^{N_P} RD_{i,j}^{\mathrm{PP}} + \sum_{j}^{N_{En}} RD_{i,j}^{\mathrm{PS}} = 1 \qquad (6)$$

Using the reciprocity properties of RD, Eq. (3) and Eq. (4) can be rewritten as

$$Q_{P,j} = \sum_{i=1}^{N_P} A_i \varepsilon_i RD_{i,j}^{\mathrm{PP}} \sigma T_i^4 = \sum_{i=1}^{N_P} A_j \varepsilon_j RD_{j,i}^{\mathrm{PP}} \sigma T_i^4 \qquad (7)$$

$$Q_{En,j} = \sum_{i=1}^{N_{En}} A_i \varepsilon_i \sigma T_i^4 RD_{i,j}^{\mathrm{SP}} = \sum_{i=1}^{N_{En}} A_j \varepsilon_j RD_{j,i}^{\mathrm{PS}} \sigma T_i^4 \qquad (8)$$

By substitution of Eq. (7) and (8) to Eq. (1), the explicit form of the particle scale energy balance is obtained as follows.

$$C_j \frac{\partial T_j}{\partial t} = \sum_{i=1}^{N_P} A_j \varepsilon_j RD_{j,i}^{\mathrm{PP}} \sigma T_i^4 + \sum_{i=1}^{N_{En}} A_j \varepsilon_j RD_{j,i}^{\mathrm{PS}} \sigma T_i^4 - A_j \varepsilon_j \sigma T_j^4 + Q_{S,j} \tag{9}$$

Based on Eq. (9), the temperature distribution in the particulate system can be solved at the particle scale with known RD and other properties of the system.

## 2.2    Continuum scale energy balance in densely particulate system

For practical engineering applications, due to huge number of particles, the calculation based on Eq. (9) will cost enormous computational effort and is usually impractical. Hence continuum theory to solve the heat transfer in the practical system is necessary. In this section, a continuum theory will be developed based on the concept of RDF, which is the generalization of the concept of RD in discrete scale (or particle scale) to continuum scale.

### 2.2.1    Definition of radiation distribution function

The generalization of the RD on the continuum scale, namely, the concept of RDF is introduced in this section. The RDF, $\widehat{RD^{\mathrm{PP}}}(\mathbf{x}_j, \mathbf{x}')$, is defined as the averaged value of the fraction of absorbed radiative power by particles in a small control volume ($\Delta V$) located at $\mathbf{x}'$ from the radiation power emitted from particle located at $\mathbf{x}_j$, as depicted in <span style="color:red">Fig. 2</span>,

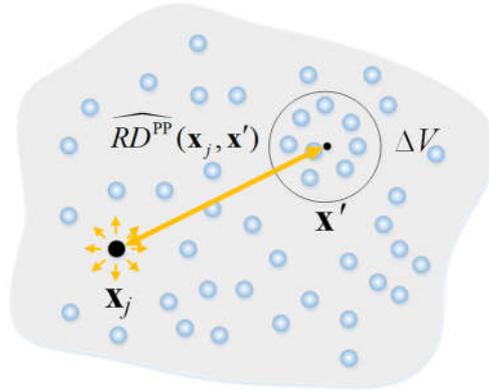

**Fig 2.**    Schematic diagram of the definition of radiation distribution function.

which can be expressed based on RD as

$$\widehat{RD^{\mathrm{PP}}}(\mathbf{x}_j, \mathbf{x}') = \frac{1}{N_{\Delta V}} \sum_{i=1}^{N_{\Delta V}} RD_{j,i}^{\mathrm{PP}} \tag{10}$$

where $N_{\Delta V} = \rho_N(\mathbf{x}')\Delta V$ denotes the number of particles in the control volume $\Delta V$, $\rho_N(\mathbf{x}')$ is the number density of particle at the position $\mathbf{x}'$ ($1/\text{m}^3$).

Assuming that the temperature of the particles in the neighborhood $\Delta V$ (position $\mathbf{x}'$) is approximately equal (namely, under local equilibrium), which is denoted as $T(\mathbf{x}')$. Based on the definition of RDF, the heating radiative power from particles located in $\Delta V$ to the particle located at $\mathbf{x}_j$ can be written as

$$
\begin{aligned}
\Delta Q_{\text{P},j} &= \sum_{i=1}^{N_{\Delta V}} A_j \varepsilon_j RD_{j,i}^{\text{PP}} \sigma T_i^4 \\
&\doteq \sigma T^4(\mathbf{x}') A_j \varepsilon_j \sum_{i=1}^{N_{\Delta V}} RD_{j,i}^{\text{PP}} \\
&= \sigma T^4(\mathbf{x}') A_j \varepsilon_j \widehat{RD^{\text{PP}}}(\mathbf{x}_j, \mathbf{x}') \rho_N(\mathbf{x}') \Delta V
\end{aligned}
\tag{11}
$$

Thus, the total radiative heating power from the particle in the entire domain to the particle located at $\mathbf{x}_j$ can be calculated from,

$$
Q_{P,j} = \int_V dQ_{\text{P},j} = \int_V A_j \varepsilon_j \widehat{RD^{\text{PP}}}(\mathbf{x}_j, \mathbf{x}') \sigma T^4(\mathbf{x}') \rho_N(\mathbf{x}') dV'
\tag{12}
$$

When the RD is varying slowly (or smoothly) with particle separation distance, the average of RD in the local volume, i.e., the RDF, is approximately equal to the interpolation or fitting function of the RD. Hence $\widehat{RD^{\text{PP}}}(\mathbf{x}_j, \mathbf{x}')$ can be determined by interpolation or fitting from the discrete scale RD ($RD_{j,i}^{\text{PP}}$) data points, which is adopt in the following analysis.

### 2.2.2 Continuum form of energy balance equation

For a differential surface element $dS'$ of enclosure wall located at $\mathbf{x}'$, the radiation power contributed to the particle located at $\mathbf{x}_j$ is

$$
dQ_{En,j} = A_j \varepsilon_j \sigma T^4(\mathbf{x}') RD^{\text{PdS}}(\mathbf{x}_j, \mathbf{x}') dS'
\tag{13}
$$

where $RD^{\text{PdS}}(\mathbf{x}_j, \mathbf{x}')$ denotes the differential RD from the particle located at $\mathbf{x}_j$ to the differential surface element at $\mathbf{x}'$ of the enclosure wall ($1/\text{m}^2$). By definition, $RD_{i,j}^{\text{PS}} = \int_S RD_i^{\text{PdS}}(\mathbf{x}')dS'$. The radiation power from the entire enclosure wall to the particle located at $\mathbf{x}_j$ is written as

$$Q_{En,j} = \oint_S dQ_{En,j} = \oint_S A_j \varepsilon_j \sigma T^4(\mathbf{x}') RD^{\mathrm{PdS}}(\mathbf{x}_j, \mathbf{x}') dS' \qquad (14)$$

With the aids of Eq. (12) and Eq. (14), the energy balance equation Eq. (1) can be rewritten as

$$
\begin{aligned}
C_j \frac{\partial T(\mathbf{x}_j)}{\partial t} = &\int_V A_j \varepsilon_j \widehat{RD^{\mathrm{PP}}}(\mathbf{x}_j, \mathbf{x}') \sigma T^4(\mathbf{x}') \rho_N(\mathbf{x}') dV' \\
&+ \oint_S A_j \varepsilon_j RD^{\mathrm{PdS}}(\mathbf{x}_j, \mathbf{x}') \sigma T^4(\mathbf{x}') dS' - A\varepsilon \sigma T^4(\mathbf{x}_j) + Q_{S,j}
\end{aligned}
\qquad (15)
$$

For further derivation, an effective occupation volume (or cell volume, $V_{\mathrm{cell},j}$) of particle located at $\mathbf{x}_j$ needs to be introduced, which can be understood as the average volume of each particle in the system. Divides both sides of Eq. (15) by $V_{\mathrm{cell},j}$, it yields

$$
\begin{aligned}
\frac{C_j}{V_{\mathrm{cell},j}} \frac{\partial T(\mathbf{x}_j)}{\partial t} = &\frac{A_j \varepsilon_j}{V_{\mathrm{cell},j}} \left[ \int_V \widehat{RD^{\mathrm{PP}}}(\mathbf{x}_j, \mathbf{x}') \sigma T^4(\mathbf{x}') \rho_N(\mathbf{x}') dV' \right. \\
&\left. + \oint_S RD^{\mathrm{PdS}}(\mathbf{x}_j, \mathbf{x}') \sigma T^4(\mathbf{x}') dS' - \sigma T^4(\mathbf{x}_j) \right] + \frac{Q_{S,j}}{V_{\mathrm{cell},j}}
\end{aligned}
\qquad (16)
$$

The above equation can be rewritten as (here the subscript $j$ is omitted for generality),

$$
\begin{aligned}
\rho_{\mathrm{eff}} c_p \frac{\partial T(\mathbf{x})}{\partial t} = &\varepsilon_V \int_V \widehat{RD^{\mathrm{PP}}}(\mathbf{x}, \mathbf{x}') \sigma T^4(\mathbf{x}') \rho_N(\mathbf{x}') dV' \\
&+ \varepsilon_V \oint_S RD^{\mathrm{PdS}}(\mathbf{x}, \mathbf{x}') \sigma T^4(\mathbf{x}') dS' - \varepsilon_V \sigma T^4(\mathbf{x}) + Q_{S,\mathrm{eff}}
\end{aligned}
\qquad (17)
$$

where $\rho_{\mathrm{eff}} = \dfrac{\rho V_P}{V_{\mathrm{cell}}}$ is the effective density (kg/m³), $Q_{S,\mathrm{eff}} = \dfrac{Q_S}{V_{\mathrm{cell}}}$ is the effective volumetric heat source (W/m³), $\varepsilon_V = \dfrac{A\varepsilon}{V_{\mathrm{cell}}}$ is called here volumetric emissivity (1/m), which characterizes the emission capability per unit volume of the media. Note that Eq. (17) is the major result of this paper, which is the continuum scale energy balance equation for particulate system. It is noted that this equation may also be applied to general porous media since the general applicability of the RD and hence RDF.

If the contribution of enclosure wall is neglected, Eq. (17) simplifies to

$$\rho_{\mathrm{eff}} c \frac{\partial T(\mathbf{x})}{\partial t} = \varepsilon_V \int_V \widehat{RD^{\mathrm{PP}}}(\mathbf{x}, \mathbf{x}') \sigma T^4(\mathbf{x}') \rho_N(\mathbf{x}') dV' - \varepsilon_V \sigma T^4(\mathbf{x}) + Q_{S,\mathrm{eff}} \qquad \mathbf{x} \in V \qquad (18)$$

In order to solve Eq. (18), appropriate boundary condition needs to be established. Since Eq. (18)

is an integral equation, the given of boundary condition is different than the differential equation, such as the heat diffusion equation. The given of boundary for Eq. (18) is discussed in the following. The whole computational domain $V$ is divided into two parts, 1) the boundary region $V_{\text{Bnd}}$ and 2) the interior region $V_{\text{Int}}$, i.e. $V = V_{\text{Int}} \cup V_{\text{Bnd}}$, as shown in Fig. 3.

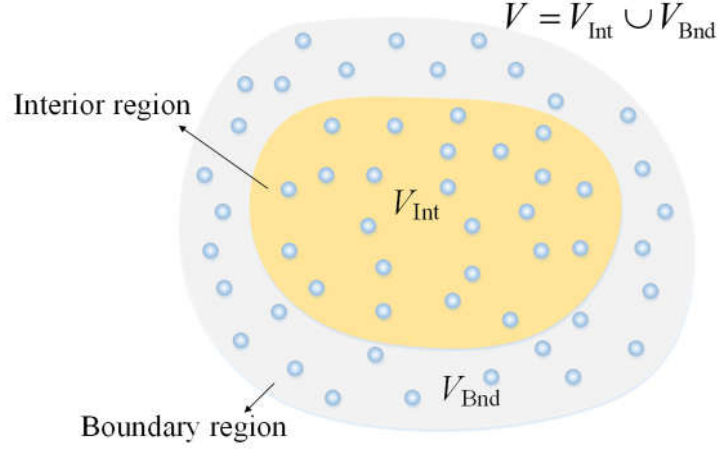

**Fig 3.** Schematic on definition of boundary condition for the continuum scale governing equation. The computational domain is divided into boundary region and interior region.

In the boundary region $V_{\text{Bnd}}$, the temperature is assumed known, given as $T_{\text{Bnd}}(\mathbf{x})$, while for the interior region $V_{\text{Int}}$, the temperature field is to be solved, hence

$$\int_V \widehat{RD^{\text{pp}}}(\mathbf{x}, \mathbf{x}') T^4(\mathbf{x}') \rho_N(\mathbf{x}') \mathrm{d}V'$$
$$= \int_{V_{\text{Bnd}}} \widehat{RD^{\text{pp}}}(\mathbf{x}, \mathbf{x}') T_{\text{Bnd}}^4(\mathbf{x}') \rho_N(\mathbf{x}') \mathrm{d}V' + \int_{V_{\text{Int}}} \widehat{RD^{\text{pp}}}(\mathbf{x}, \mathbf{x}') T^4(\mathbf{x}') \rho_N(\mathbf{x}') \mathrm{d}V'$$

(19)

Eq. (18) can be rewritten as

$$\rho_{\text{eff}} c \frac{\partial T(\mathbf{x})}{\partial t} = \varepsilon_V \sigma \int_{V_{\text{Int}}} \widehat{RD^{\text{pp}}}(\mathbf{x}, \mathbf{x}') T^4(\mathbf{x}') \rho_N(\mathbf{x}') \mathrm{d}V' - \varepsilon_V \sigma T^4(\mathbf{x}) + Q_{S,\text{eff}}$$
$$+ \varepsilon_V \sigma \int_{V_{\text{Bnd}}} \widehat{RD^{\text{pp}}}(\mathbf{x}, \mathbf{x}') T_{\text{Bnd}}^4(\mathbf{x}') \rho_N(\mathbf{x}') \mathrm{d}V' \qquad \mathbf{x} \in V_{\text{Int}}$$

(20)

It should be noted that since the RDF decays rapidly with distance (as shown in Section 4), the influence of the boundary region located far from the interior region on heat transfer within the interior region is negligible.

For steady state problem, Eq. (20) is simplified as follows,

$$T^4(\mathbf{x}) - \int_{V_{\text{Int}}} \widehat{RD^{\text{PP}}}(\mathbf{x},\mathbf{x}')T^4(\mathbf{x}')\rho_N(\mathbf{x}')\mathrm{d}V'$$

$$= \frac{Q_{S,\text{eff}}}{\varepsilon_V \sigma} + \int_{V_{\text{Bnd}}} \widehat{RD^{\text{PP}}}(\mathbf{x},\mathbf{x}')T_{\text{Bnd}}^4(\mathbf{x}')\rho_N(\mathbf{x}')\mathrm{d}V' \qquad \mathbf{x} \in V_{\text{Int}} \qquad (21)$$

If there is no heat source, Eq. (21) can be further simplified as

$$T^4(\mathbf{x}) - \int_{V_{\text{Int}}} \widehat{RD^{\text{PP}}}(\mathbf{x},\mathbf{x}')T^4(\mathbf{x}')\rho_N(\mathbf{x}')\mathrm{d}V'$$

$$= \int_{V_{\text{Bnd}}} \widehat{RD^{\text{PP}}}(\mathbf{x},\mathbf{x}')T_{\text{Bnd}}^4(\mathbf{x}')\rho_N(\mathbf{x}')\mathrm{d}V' \qquad \mathbf{x} \in V_{\text{Int}} \qquad (22)$$

Corresponding to the discrete form of conservation property (Eq. (6)) of RD, the conservation property of RDF can be written as

$$\int_V \widehat{RD^{\text{PP}}}(\mathbf{x}_j,\mathbf{x}')\rho_N(\mathbf{x}')\mathrm{d}V' + \oint_S RD^{\text{PdS}}(\mathbf{x}_j,\mathbf{x}')\mathrm{d}S' = 1 \qquad (23)$$

If the effect of enclosure wall can be omitted, the conservation property reduces to

$$\int_V \widehat{RD^{\text{PP}}}(\mathbf{x}_j,\mathbf{x}')\rho_N(\mathbf{x}')\mathrm{d}V' = 1 \qquad (24)$$

For one-dimensional particulate system, the cell volume is in unit of length, denoted as $L_{\text{cell}}$, i.e. $V_{\text{cell}} \triangleq L_{\text{cell}}$, hence $\varepsilon_V = \frac{A\varepsilon}{V_{\text{cell}}} \triangleq \frac{A\varepsilon}{L_{\text{cell}}}$ and $\rho_{\text{eff}} = \frac{\rho V_P}{V_{\text{cell}}} \triangleq \frac{\rho V_P}{L_{\text{cell}}}$. In this case, $\rho_N$ is the number of particles per unit length (1/m) and $Q_{S,\text{eff}}$ denotes heating power per unit length (W/m). For two-dimensional system, the cell volume is in unit of area, denoted as $A_{\text{cell}}$, i.e. $V_{\text{cell}} \triangleq A_{\text{cell}}$, hence $\varepsilon_V = \frac{A\varepsilon}{V_{\text{cell}}} \triangleq \frac{A\varepsilon}{A_{\text{cell}}}$ and $\rho_{\text{eff}} = \frac{\rho V_P}{V_{\text{cell}}} \triangleq \frac{\rho V_P}{A_{\text{cell}}}$. $\rho_N$ is the number of particles per unit area (1/m²), and $Q_{S,\text{eff}}$ denotes heating power per unit area (W/m²).

### 2.2.3    Treatment of singular point

Because of the interaction among particles, a part of the emitted power by particle $j$ may be absorbed by itself, i.e. $RD_{j,j}^{\text{PP}}$ is not zero. Sometimes, $RD_{j,j}^{\text{PP}}$ is significantly different from RD of other position, as shown in Fig. 4, which will cause difficulty in the obtaining of RDF from interpolation or fitting, resulting in complex function form that are difficult for accurate numerical integration. In this case, a special treatment for the self-absorption term is required. The $RD_{j,j}^{\mathrm{pp}}$ can be regarded as a singular point and the RDF can be described as a combination of a delta function (representing the self-absorption term) and a smooth function. A special treatment for the self-absorption term $RD_{j,j}^{\mathrm{pp}}$ is presented here.

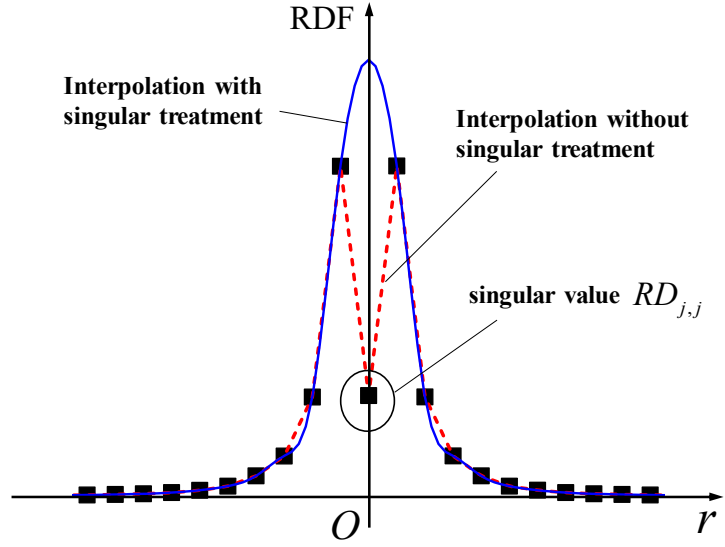

**Fig 4.** Schematic diagram of RD singular point. RDF is obtained by interpolation.

By separating the self-absorption term, the conservation property of RD at the discrete scale can be written as

$$\sum_{i \neq j}^{N_P} RD_{j,i}^{\mathrm{pp}} + RD_{j,j}^{\mathrm{pp}} + \sum_i^{N_{En}} RD_{j,i}^{\mathrm{ps}} = 1 \tag{25}$$

By treating the whole RDF as a superposition of a delta function (representing the self-absorption term) and a slowly varying function, the conservation property of RDF (Eq. (23)) can be rewritten as

$$\int_{V/\mathbf{x}_j} \widehat{RD^{\mathrm{pp}}}(\mathbf{x}_j, \mathbf{x}') \rho_N(\mathbf{x}') \mathrm{d}V' + RD_{j,j}^{\mathrm{pp}} + \oint_S RD^{\mathrm{PdS}}(\mathbf{x}_j, \mathbf{x}') \mathrm{d}S' = 1 \tag{26}$$

where $V/\mathbf{x}_j$ represents the computational domain $V$ without the singular point at $\mathbf{x}_j$. In this case, $Q_{\mathrm{P},j}$ can be rewritten as

$$Q_{P,j} = \int_V \mathrm{d}Q_{\mathrm{P},j} = \int_{V/\mathbf{x}_j} A_j \varepsilon_j \widehat{RD^{\mathrm{PP}}}(\mathbf{x}_j, \mathbf{x}') \sigma T^4(\mathbf{x}') \rho_N(\mathbf{x}') \mathrm{d}V' + A_j \varepsilon_j \sigma T^4(\mathbf{x}_j) RD_{j,j}^{\mathrm{PP}} \qquad (27)$$

When the effect of enclosure wall on heat transfer is not considered, the conservation property (Eq. (26)) becomes

$$\int_{V/\mathbf{x}_j} \widehat{RD^{\mathrm{PP}}}(\mathbf{x}_j, \mathbf{x}') \rho_N(\mathbf{x}') \mathrm{d}V' + RD_{j,j}^{\mathrm{PP}} = 1 \qquad (28)$$

With the singularity treatment, the energy balance equation at the continuum scale (Eq. (17)) can be rewritten as

$$\rho_{\mathrm{eff}} c \frac{\partial T(\mathbf{x})}{\partial t} = \varepsilon_V \int_{V/\mathbf{x}} \widehat{RD^{\mathrm{PP}}}(\mathbf{x}, \mathbf{x}') \sigma T^4(\mathbf{x}') \rho_N(\mathbf{x}') \mathrm{d}V' + \varepsilon_V \left( RD_{j,j}^{\mathrm{PP}} - 1 \right) \sigma T^4(\mathbf{x}) + Q_{S,\mathrm{eff}} \qquad \mathbf{x} \in V \quad (29)$$

The establishment of boundary condition for Eq. (29) is same as that for Eq. (17).

## 2.3 Numerical discretization and solution

A particulate system composed of identical particles is considered. The particles are evenly distributed, opaque and isothermal. The effect of enclosure wall is omitted. According to Eq. (22), the one-dimensional continuum form of energy balance equation is given as

$$T^4(x) - \rho_N \int_{V_{\mathrm{Int}}} \widehat{RD^{\mathrm{PP}}}(x, x') T^4(x') \mathrm{d}x' = \rho_N \int_{V_{\mathrm{Bnd}}} \widehat{RD^{\mathrm{PP}}}(x, x') T_{\mathrm{Bnd}}^4(x') \mathrm{d}x' \qquad x \in V_{\mathrm{Int}} \qquad (30)$$

The one-dimensional domain is discretized into many intervals with spacing of $\Delta x$, as shown in Fig. 5(c). The size of the exterior region must be large enough to ensure that the energy emitted from the interior region to enclosure wall is negligible, which can be determined based on the decaying properties of the RDF, as shown in Fig. 5(a).

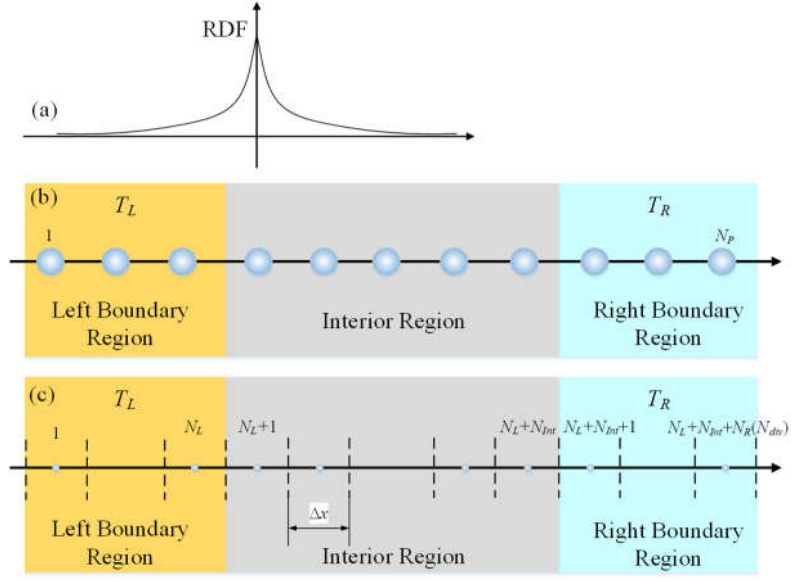

**Fig 5.**     Schematic diagram of the division of one dimensional computational domain. (a) RDF distribution, (b)

particle scale computational domain, (c) continuum scale computational domain.

By applying numerical integration over the intervals, Eq. (30) can be written in discretized form

as

$$
\begin{aligned}
& T_j^4 - \rho_N \sum_{i=N_L+1}^{N_L+N_{\text{Int}}} \widehat{RD^{\text{PP}}}(x_j, x_i) T_i^4 \Delta x \\
& = \rho_N \left[ \sum_{i=1}^{N_L} \widehat{RD^{\text{PP}}}(x_j, x_i) T_L^4 \Delta x + \sum_{i=N_L+N_{\text{Int}}+1}^{N_{\text{div}}} \widehat{RD^{\text{PP}}}(x_j, x_i) T_R^4 \Delta x \right]
\end{aligned}
\qquad j = N_L+1,..., N_L + N_{Int} \quad (31)
$$

where $T_L$ and $T_R$ are the given temperature of the left and right exterior regions, respectively, $N_{\text{div}}$

denotes the total number of intervals. In order to make the discretized RDF in Eq. (31) preserve the conservation property (Eq.(24)), it is modified or normalized as follows.

$$
\widehat{RD^{\text{PP}}}(x_j, x_i) = \frac{\widehat{RD^{\text{PP}}}(x_j, x_i)}{\sum_{i=1}^{N_{\text{div}}} \widehat{RD^{\text{PP}}}(x_j, x_i) \rho_N \Delta x} \qquad j = N_L+1,..., N_L + N_{Int} \qquad (32)
$$

For the steady state problem with singularity treatment, according to the Eq. (29), the one-

dimensional continuum form of energy balance equation becomes

$$
\begin{aligned}
\left(1 - RD_{j,j}^{\text{PP}}\right) T^4(x) = & \int_{V_{\text{Int}}/x} \widehat{RD^{\text{PP}}}(x, x') T^4(x') \rho_N(x') \mathrm{d}x' \\
& + \int_{V_{\text{Bnd}}} \widehat{RD^{\text{PP}}}(x, x') T_{\text{Bnd}}^4(x') \rho_N(x') \mathrm{d}x'
\end{aligned}
\qquad x \in V_{\text{Int}} \qquad (33)
$$

Further, Eq. (33) can be discretized as

$$
\begin{aligned}
T_j^4 &- \frac{\rho_N}{1-RD_{j,j}^{\mathrm{PP}}} \sum_{i=N_L+1}^{N_L+N_{\mathrm{Int}}} \widehat{RD^{\mathrm{PP}}}(x_j,x_i)T_i^4 \Delta x \\
&= \frac{\rho_N}{1-RD_{j,j}^{\mathrm{PP}}} \left[ \sum_{i=1}^{N_L} \widehat{RD^{\mathrm{PP}}}(x_j,x_i)T_L^4 \Delta x + \sum_{i=N_L+N_{\mathrm{Int}}+1}^{N_{\mathrm{div}}} \widehat{RD^{\mathrm{PP}}}(x_j,x_i)T_R^4 \Delta x \right]
\end{aligned} \quad i=N_L+1,...,N_L+N_{Int} \quad (34)
$$

In order to guarantee the conservation property of RDF (Eq. (28)), the discrete form of radiation distribution function is modified or normalized as follows.

$$
\widehat{RD^{\mathrm{PP}}}(x_j,x_i) = \frac{\widehat{RD^{\mathrm{PP}}}(x_j,x_i)}{\sum\limits_{i=1}^{N_{\mathrm{div}}} \widehat{RD^{\mathrm{PP}}}(x_j,x_i)\rho_N \Delta x}(1-RD_{j,j}^{\mathrm{PP}}) \qquad i=N_L+1,...,N_L+N_{Int} \quad (35)
$$

## 3 Numerical model for determining RDF

From the derivation presented in Section 2, the temperature distribution can be solved using Eq. (9) (at particle scale) with RD as input physical property of the system, or using Eq. (17) (under continuum approach) with RDF as input physical property. Hence, to determine the RDF is essential to solve the system temperature distribution based on the continuum approach. Here in this section, numerical model to calculate the RD or RDF is presented. Since by definition, RDF can be considered the interpolation or fitting function of the RD, the RDF and RD can be considered equivalent to large extent in physics. Hence in the following, we will not make a clear distinction between the RDF and RD.

### 3.1 Monte Carlo ray tracing algorithm

Detailed radiative heat transfer simulation is required to determine the RDF. In this work, the Monte Carlo ray tracing (MCRT) algorithm is applied to determine the RDF in particulate media, which fully considers the 'volume scattering' and multiple scattering effects [44-48]. During the MCRT simulation, the radiation is treated as a superposition of pencils of rays (or photon bundles), with a given quantity of energy (initially set to be unit), propagating along a straight path. A photon bundle is followed from their emission point through a series of reflections on the particle surface, a part of its energy is absorbed when it hits a particle, until the energy of the photon bundle is below a preset minimum value or it escapes from the computational domain. More details concerning this approach can be found in reference [42]. By emitting a large number of rays from the source particle and recording the number of absorbed photon bundles, the RD can be calculated by

$$RD_{i,j} = \frac{N_j}{N_{tot}}$$ (36)

where $N_j$ is the number of absorbed photon bundles by particle $j$, $N_{tot}$ refers to the total number of photon bundles emitted from the source particle $i$.

Note that the starting position $(x, y, z)$ of photon bundles must be located uniformly on the surface of source particle $i$, which is calculated from

$$
\begin{aligned}
x &= x_i + R \cdot \sqrt{1 - \xi_1^2} \cos\left(2\pi\xi_2\right) \\
y &= y_i + R \cdot \sqrt{1 - \xi_1^2} \sin\left(2\pi\xi_2\right) \\
z &= z_i + R \cdot \xi_1
\end{aligned}
$$ (37)

where $\xi_1$ and $\xi_2$ are random numbers between 0 and 1, $x_i$, $y_i$, $z_i$ represent the coordinate of the source particle $i$.

The interaction of photon bundles with optically large particles takes place at the interface and the surface type of particles have an important effect on the interaction. In this work, there are two kinds of surface reflection modes are considered, i.e., diffuse and specular reflection. For diffuse surface particles, the direction of reflection is independent of the incident angle and it can be determined by lambert's cosine law [49],

$$
\begin{aligned}
\theta &= \arccos(\sqrt{1 - \xi_1}) \\
\varphi &= 2\pi\xi_2
\end{aligned}
$$ (38)

where $\theta$ and $\varphi$ denote respectively the azimuth angle and zenith angle of the local coordinate system. While for specular surface particles, $\theta_{re} = \theta_{in}$, $\theta_{re}$ and $\theta_{in}$ denote the angle of reflection and incident, respectively. In this case, reflectivity is determined by Fresnel's law [9] and the reflected parts of energy are

$$\rho_{\parallel} = \frac{(p' - \sin\theta_{in}\tan\theta_{in})^2 + q'^2}{(p' + \sin\theta_{in}\tan\theta_{in})^2 + q'^2}\rho_{\perp}$$ (39)

$$\rho_{\perp} = \frac{(\cos\theta_{in} - p')^2 + q'^2}{(\cos\theta_{in} + p')^2 + q'^2}$$ (40)

where

$$p'^2 = \frac{1}{2}\left[\sqrt{(n^2-k^2-\sin^2\theta_i)^2+4n^2k^2}+(n^2-k^2-\sin^2\theta_{in})\right]$$
$$q'^2 = \frac{1}{2}\left[\sqrt{(n^2-k^2-\sin^2\theta_i)^2+4n^2k^2}-(n^2-k^2-\sin^2\theta_{in})\right]$$

(41)

$\rho_{\parallel}$ and $\rho_{\perp}$ denote the parallel and perpendicular polarized radiations respectively. $n$ and $k$ denote the optical constant of material of particles. The Fresnel's reflectivity for the non-polarized irradiation can be calculated from

$$\rho_R = (\rho_{\parallel}+\rho_{\perp})/2$$

(42)

In addition, the particles are assumed as opaque and Kirchhoff's law is applied,

$$\varepsilon = \alpha = 1-\rho_R$$

(43)

where $\varepsilon$, $\alpha$ and $\rho_R$ denote particle wall emissivity, absorptivity and reflectivity, respectively. Three different cases are considered, 1) SiC particles under irradiation with wavelength of 5μm and 2) 10μm, and 3) Ti particles are irradiated at a wavelength of 1μm. The optical constants used under different conditions are obtained from Ref. [50], are shown in Table 1. The variation of particle wall emissivity $\varepsilon$ with incident angle $\theta_{in}$ for different conditions are shown in Fig. 6.

Table 1.   Optical constants used under different conditions.

| Material | Wavelength(μm) | $n$ | $k$ |
|----------|----------------|-----|-----|
| SiC | 5 | 2.467 | 0.00035 |
| SiC | 10 | 1.050695 | 0.03416 |
| Ti | 1 | 3.396 | 3.242 |

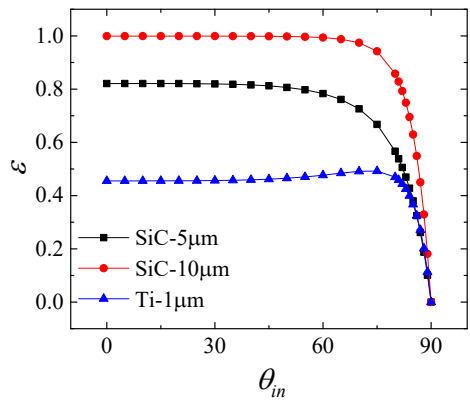

**Fig 6.**   Variation of particle wall emissivity $\varepsilon$ with incident angle $\theta_{in}$ for different cases.

## 3.2 Model validation

Considering a special case that the particles are black, the value of RD will be equal to the value of the view factor. There are exact results of view factors for two spherical particle configuration [51], which can be used for the validation of the calculation model of RDF proposed in this work. Figure 7 shows a comparison of the simulated results with exact solutions for the same geometric configuration.

Increasing the distance between two particles, the results of comparing the numerical model and exact solution [40] are shown in Fig. 7(b). It can be seen that the simulated results in both Figs. 7(a) and (b) are in good agreement with the exact solution. It can be noted that in Fig. 7(a), when the number of light reaches $5 \times 10^6$, the RD tends to converge to the exact result. The number of $5 \times 10^6$ photon bundles is used in this work, which has been verified to give converged results.

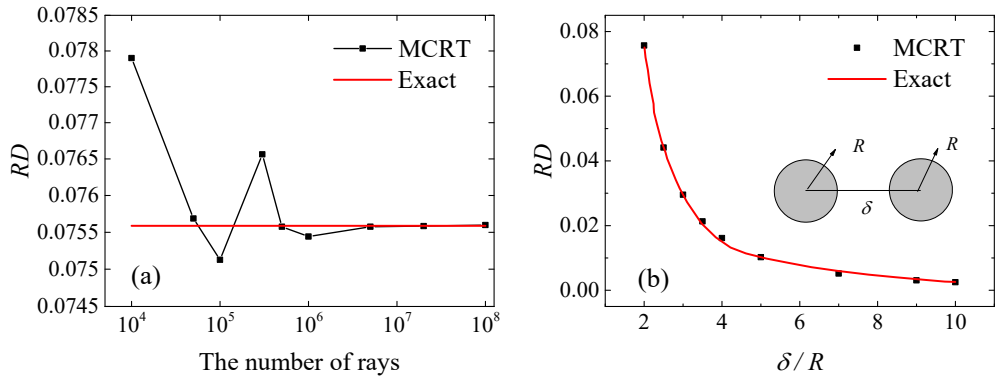

**Fig 7.** Convergence test of RD and comparison of exact and MCRT results: **(a)** convergence test for two touching particles **(b)** RD for two particles at different particle spacing $\delta$.

## 4 Results and discussions

The role of RDF in the continuum approach is similar to the radiative properties in RTE. By definition, the RDF can be determined as the interpolation or fitting function of the RD. In the following, the RDF of different arrangement of particles is studied numerically. The continuum approach is verified by the heat transfer simulation at particle scale.

In this work, two types of arrangements, i.e. randomly and regularly distribution, are considered. The influence of three main factors on the distribution of RD are analyzed, namely, the particle filling ratio $f$, the particle wall emissivity $\varepsilon$ and reflection properties (i.e. specular and diffuse). It's worth noting that the particle filling ratio and dimensionless particle spacing $\delta^*$ (center to center, normalized to $R$) are two dependent parameters, $f = \frac{4}{3}\pi(\frac{R}{\delta})^3 = \frac{4}{3}\pi\delta^{*-3}$. Hence only one is needed for the analysis. In the following, a dimensionless distance $r^*=r/R$ is used, which stands for the distance from the source particle that is normalized with the particle radius $R$.

## 4.1    RDF in randomly distributed particulate system

The RDF in randomly distributed particulate system is statistically meaningful. A 'double average' treatment is adopted in this paper to reduce the statistical error caused by random arrangement, as illustrated in Fig. 8. More specifically, the 'first average' refers to the average of particles within spherical shells centered as source particle with thickness (take the average particle spacing in this work) and the 'second average' stands for the average for different realization of random particle distribution.

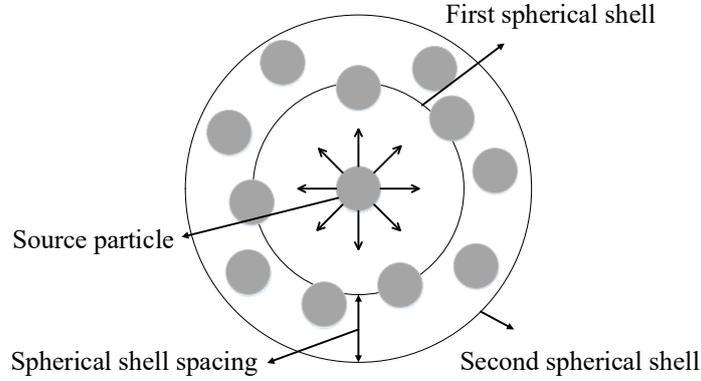

**Fig 8.**    Illustration of 'double average' treatment.

### 4.1.1    Theoretical model of the radiation distribution function

In the following, a theoretical model of RDF in the randomly distributed particulate system is presented. Generally, the farther a particle from the source particle, the fewer it will receive and absorb radiative power. Due to geometric symmetry in the randomly distributed system, the RDF can be seen as a function of the relative distance from the source particle $r = \left| \mathbf{x}_j - \mathbf{x}' \right|$. Considering a system shown in Fig. 9, the RDF at distance $r$ from the source particle can be written as

$$RDF(r) = \frac{\alpha A \, q(r)}{Q} \tag{44}$$

where $Q$ stands for the total emitted power of the volume element $j$ (corresponding to the source particle $j$ at the position $\mathbf{x}_j$), $q(r)$ and $\alpha$ denote respectively the radiative flux density and the absorptivity of the particles within the neighborhood $\varDelta V$ at $\mathbf{x}'$. $A$ is the projected area of the particles in $\varDelta V$.

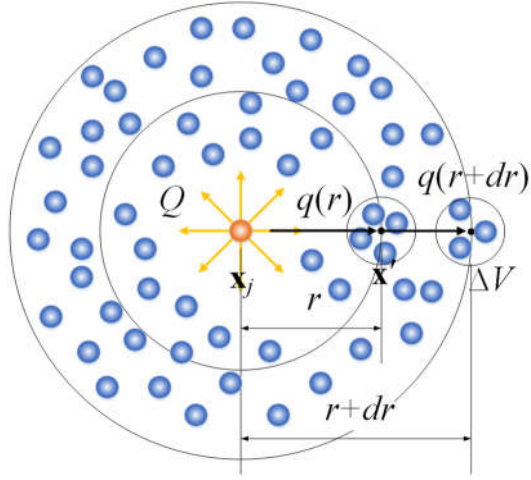

**Fig 9.**    Schematic representation of randomly distributed particulate system.

During the propagation of radiation from the volume element *j*, two kinds of effects influence $q(r)$ that need to be considered simultaneously, one is the attenuation caused by the increase of radial distance, $(\delta q)_1 = -\dfrac{2}{r}q(r)dr$, due to the expansion of spherical beam front, and the other is the attenuation caused by the absorption and scattering of particles, $(\delta q)_2 = -\beta q(r)dr$, which follows Bouguer's law [9]. So

$$dq = (\delta q)_1 + (\delta q)_2 = -\left(\beta + \frac{2}{r}\right)q(r)dr \qquad (45)$$

Consider the general solution of Eq. (45) and refer to Eq. (44),

$$RDF(r) = C\frac{\alpha A}{Q}\frac{1}{r^2}e^{-\beta r} \qquad (46)$$

namely, the RDF follows a combination of inverse square law and the exponential decay law with radial distance *r*. During the expansion of spherical beam front, there will be the obstruction of particles and Eq. (46) need to be modified. Therefore, generalize the Eq. (46) and the theoretical model of RDF of randomly distributed particle system can be given,

$$RDF(r) = \frac{\gamma}{(\alpha_0 + \alpha_1 r^1 + \alpha_2 r^2)}e^{-\beta r} \qquad (47)$$

in which $\alpha_0$, $\alpha_1$, $\alpha_2$, $\beta$ and $\gamma$ are undetermined coefficient. Note that $\beta$ is retrieved extinction coefficient. The performance of fitting using Eq. (47) is significantly improved compared with the model presented in Ref. [52],

$$RDF(r) = \frac{C}{r^2}e^{-\beta r} \qquad (48)$$

Note that the mode is not limited to spherical particles as indicated from the derivation.

### 4.1.2 Particles of diffuse surface

The RDF as a function of $r^*$ for different particle filling ratio $f$ are shown in Fig. 10. Particle wall emissivity $\varepsilon$ is set as 1. It can be seen that the RDF decay rapidly with increasing radial distance $r^*$ and the decaying rate increases at larger $f$. It's attributed to the attenuation caused by the increase of radial distance and the absorption of particles. Moreover, the effect of absorption of particles increases with increasing $f$.

Following the derivation presented in Section 4.1.1, the variation trend of the RDF can be analyzed by comparing with the inverse square law and the exponential decay law. As shown in Fig. 10, the data points of the simulated RDs are well between the fitted results of the inverse square law ($a/(r^*)^2$) and the exponential decay law ($be^{-cr^*}$). For low $f$, the simulated result is closer to the inverse square law. While for high $f$, it is closer to the exponential decay law.

The theoretical model of RDF for random particle arrangement is verified by fitting the simulated results. It's worth noting that the fitting curve, calculated by Eq. (47), agrees very well with the data points of MCRT method for different $f$. The fitted parameters based on Eq. (47) for different $f$ are reported in Table 2. Note that the retrieved extinction coefficient $\beta$ increases with the increase $f$.

Table 2. Fitted parameters of the theoretical model of RDF (Eq. (47)) for different filling ratio, the particle wall emissivity is 1.

| $f$ | 0.048 | 0.194 | 0.291 |
|---|---|---|---|
| $\alpha_0$ | 2.854 | 2.705 | 2.395 |
| $\alpha_1$ | 0 | 0 | 0 |
| $\alpha_2$ | 1 | 1 | 1 |
| $\beta$ | 0.039 | 0.189 | 0.305 |
| $\gamma$ | 0.295 | 0.456 | 0.573 |

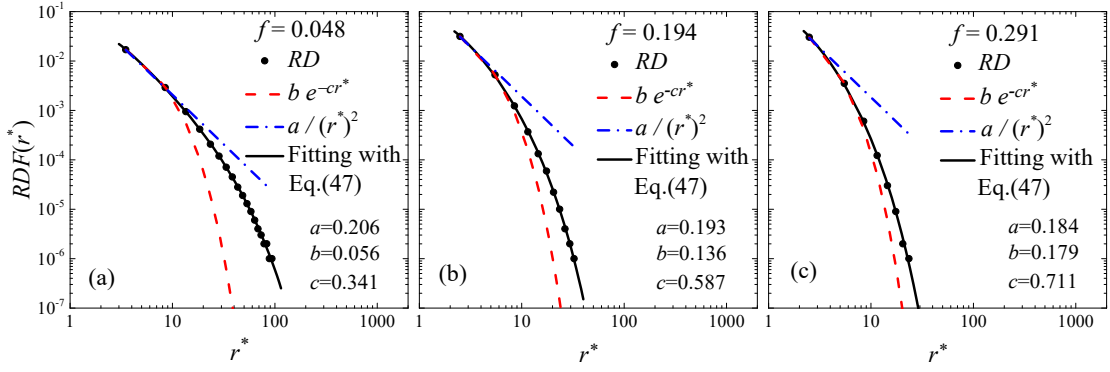

**Fig 10.** Variation of the RDF with radial distance for different particle filling ratio $f$ with a fixed particle wall emissivity $\varepsilon$ of 1: **(a)** $f$=0.048, **(b)** $f$=0.194, **(c)** $f$=0.291. The figure shows the parameters of reference.

The influence of particle wall emissivity $\varepsilon$ on the distribution of RDF is shown in Fig. 11. The theoretical model predicted RDF (fitting the numerical data based on Eq. (47)) are presented in Fig. 11(a), and that of the previous prediction model (Eq. (48)) are presented in Fig. 11(b). To ensure that the results are not affected by the finite computational domain, the energy escaped from the computational domain must be small enough. The particle filling ratio $f$ is fixed as 0.194. As shown, the decaying rate of RDF decreases at smaller $\varepsilon$, which is due to the enhancement of multiple scattering among particles. The fitted RDF using theoretical model is in good agreement with result of MCRT simulation for different $\varepsilon$. Moreover, compared with Eq. (48), the fitting function using Eq. (47) is more accurate as shown in Fig. 11. The fitted parameters based on Eq. (47) for different $\varepsilon$ are reported in Table 3. The retrieved extinction coefficient $\beta$ increases with the increase of $\varepsilon$.

Table 3. Fitted parameters of the theoretical model (Eq. (47)) for different particle wall emissivity, the particle filling ratio is 0.194.

| $\varepsilon$ | 0.4 | 0.8 | 1 |
|---|---|---|---|
| $\alpha_0$ | 9.168 | 7.989 | 2.704 |
| $\alpha_1$ | 0 | 0 | 0 |
| $\alpha_2$ | 1 | 1 | 1 |
| $\beta$ | 0.154 | 0.188 | 0.189 |
| $\gamma$ | 0.44 | 0.574 | 0.456 |

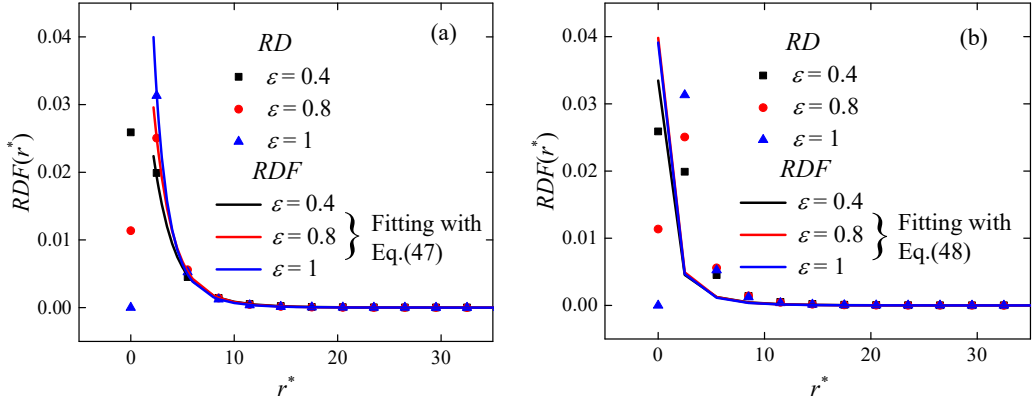

**Fig 11.** Variation of the RDF with radial distance for different particle wall emissivity $\varepsilon$ with a fixed particle filling ratio $f$ of 0.194. **(a)** fitting with the theoretical model in this work (Eq. (47)); **(b)** fitting with the theoretical model of Ref [52] (Eq. (48)).

### 4.1.3   Particles of specular surface

For particles with specular surface, the direction of reflection highly depends on the incident direction. The $\varepsilon$ (or $\alpha$) is determined using Fresnel's law as illustrated in Section 3.1. The influence of particle filling ratio $f$ and particle wall emissivity $\varepsilon$ on the distribution of RDF are shown in Figs. 12 and 13, respectively. The fitted parameters using the theoretical model are presented in Tables 4 and 5 respectively. The trend of RDF is similar to the previous case of diffuse surface particles. As shown, when increase $f$, the decay rate of RDF increase as well. The fitting function of RDFs with theoretical model (Eq. (47)) agree very well with the results of MCRT simulation. Generally, for all cases in a randomly distributed system, the fitting function of RDFs using the theoretical model (Eq. (47)) are in close match with the results from MCRT simulation, which demonstrates the good applicability of the presented theoretical RDF model.

Table 4.   Fitted parameters of the theoretical RDF model (Eq. (47)) for different filling ratio, the particle is made of SiC, and the wavelength is 10μm.

| $f$ | 0.048 | 0.194 | 0.291 |
|---|---|---|---|
| $\alpha_0$ | 3.317 | 3.089 | 7.248 |
| $\alpha_1$ | 0 | 0 | 0 |
| $\alpha_2$ | 1 | 1 | 1 |
| $\beta$ | 0.039 | 0.197 | 0.343 |
| $\gamma$ | 0.262 | 0.455 | 0.918 |

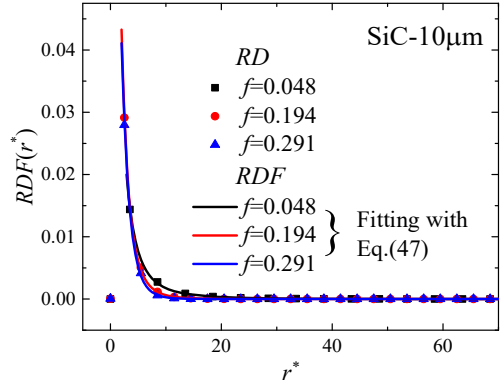

**Fig 12.** Variation of the RDF with radial distance for different filling ratio,

the particle is SiC and wave length is 10μm.

Table 5. Fitted parameters of the theoretical model (Eq. (47)) for different conditions,

particle filling ratio is 0.291.

| material | SiC-5μm | SiC-10μm | Ti-1μm |
|----------|---------|----------|--------|
| $\alpha_0$ | 4.8 | 7.248 | 10.63 |
| $\alpha_1$ | 0 | 0 | 0 |
| $\alpha_2$ | 1 | 1 | 1 |
| $\beta$ | 0.313 | 0.343 | 0.274 |
| $\gamma$ | 0.678 | 0.919 | 0.713 |

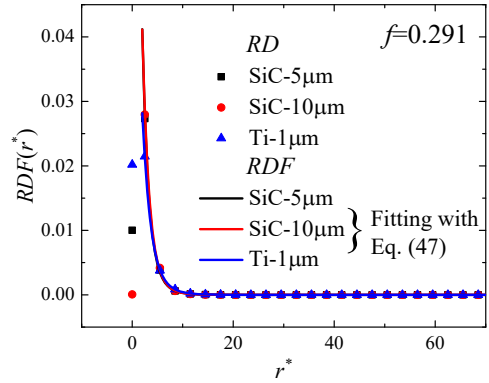

**Fig 13.** Variation of the RDF with distance parameter $r^*$ for different conditions

with a fixed particle filling ratio of 0.291.

## 4.2 RDF in regularly distributed particulate system

Comparing with statistically isotropic RDF in the randomly packed system, the RDF distribution in a regularly distributed particulate system (like crystal structure) will show significant anisotropy, as depicted in Fig. 14.

For the particle arrangement considered, there are three characteristic directions, namely,

directions 1, 2 and 3 as shown in Fig. 14. The RDF distribution along those three characteristic directions will be different. For the regularly distributed particle system, the dimensionless particle spacing (the lengths of cell edges) $\delta^* = (\delta x^*, \delta y^*, \delta z^*)$ instead of $f$ is used for characterization the system.

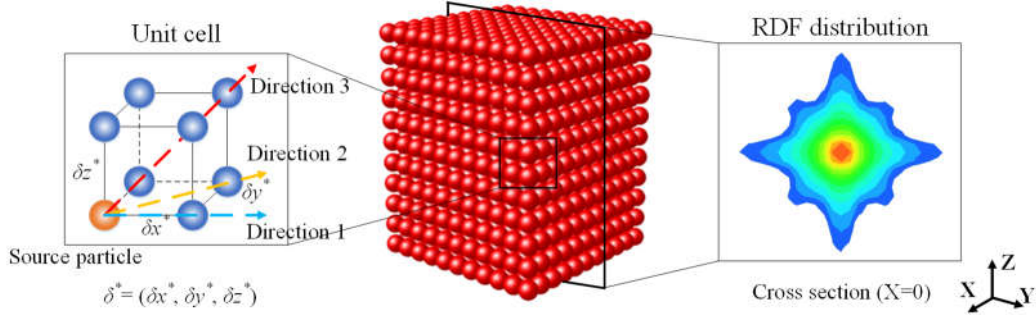

**Fig 14.** Schematic diagram of a regularly distributed particulate system.

The simple cubic structure ($\delta x^* = \delta y^* = \delta z^* = \delta^*$) is considered. The RDF distribution at cross section $X = 0$, for the case of $\varepsilon = 0.4$ and $\delta^* = 2$, is shown in Fig. 15. The results reveal anisotropic characteristics of the RDF distribution. Figure. 16 depicts the distribution of RDF along three characteristic directions for different $\delta^*$. The relationship between $\delta^*$ and $f$ for the cubic lattice arrangement is shown in Table 6. Three different particle spacing $\delta^*$ are considered, namely 2, 2.5 and 3. It's shown that in all directions, the decaying rate of RDF decreases at larger $\delta^*$, since the absorption of particles increases with increasing $f$. Note that the variation of RDF along three directions changes a lot.

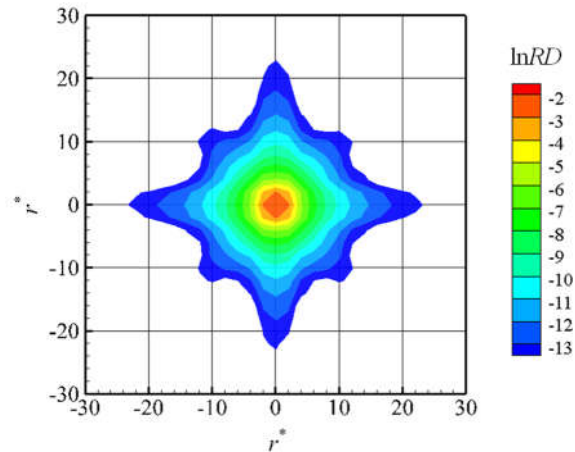

**Fig 15.** RDF distribution of cross section X=0 for regular arrangement of particles with diffuse surface.

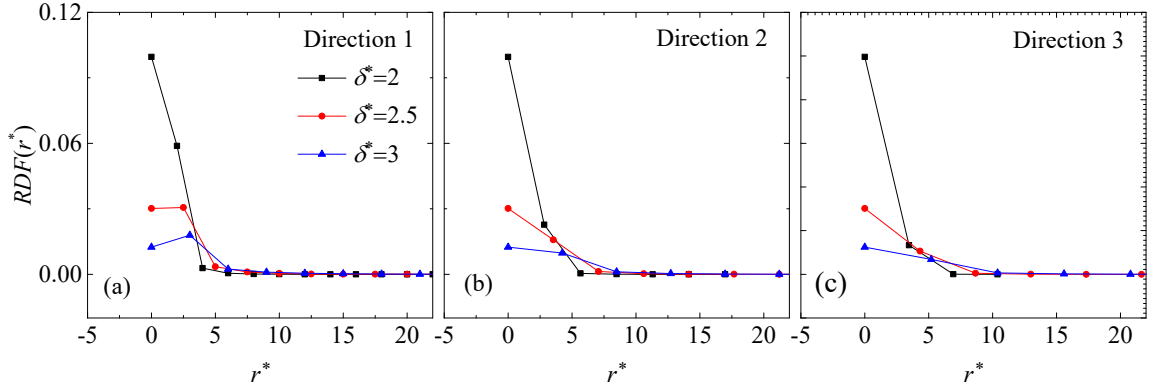

**Fig 16.** Variation of the RDF along three directions at different particle spacing, particle wall emissivity is 0.4: **(a)** direction 1, **(b)** direction 2, **(c)** direction 3.

Table 6. Corresponding relation between the dimensionless particle spacing and particle filling ratio.

| $\delta^*$ | 2 | 2.5 | 3 |
|---|---|---|---|
| $f$ | 0.524 | 0.268 | 0.155 |

The variation of the RDF along three directions for different particle wall emissivity is shown in Fig. 17. The particle spacing $\delta^*$ is set as 3. When $\varepsilon$ is small, multiple reflection among particles is prominent, while the absorption of particles is weak. When $\varepsilon$ increases, the decaying rate of RDF increases, indicating the attenuation by particles becomes stronger. The RDF distributions along three characteristic directions when the particle spacing $\delta x^*$, $\delta y^*$ and $\delta z^*$ are not equal are shown in Fig. 18. It's shown that when one or two of cell edges have been lengthened, the decay rate of RDF decreases due to the increase of particle clearance.

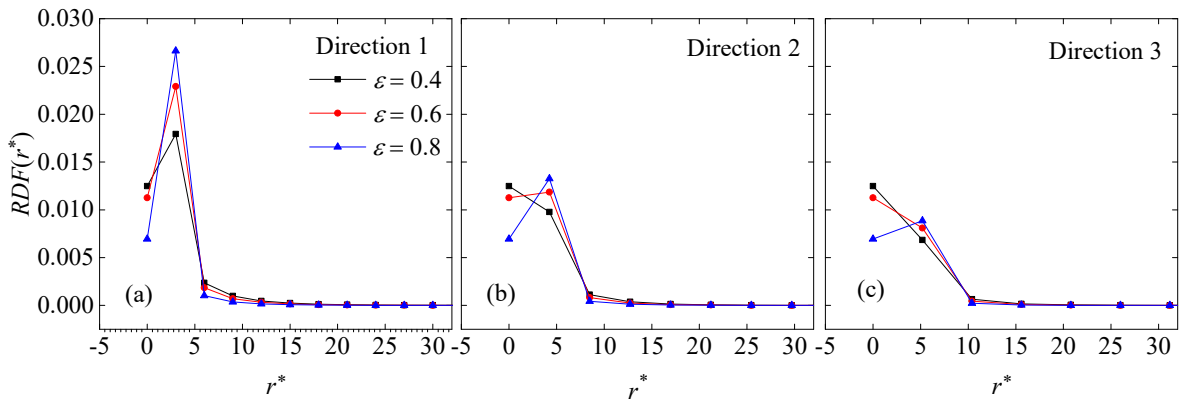

**Fig 17.** Variation of the RDF along three directions for different particle wall emissivity ε, the particle spacing $\delta^*$=3: **(a)** direction 1, **(b)** direction 2, **(c)** direction 3.

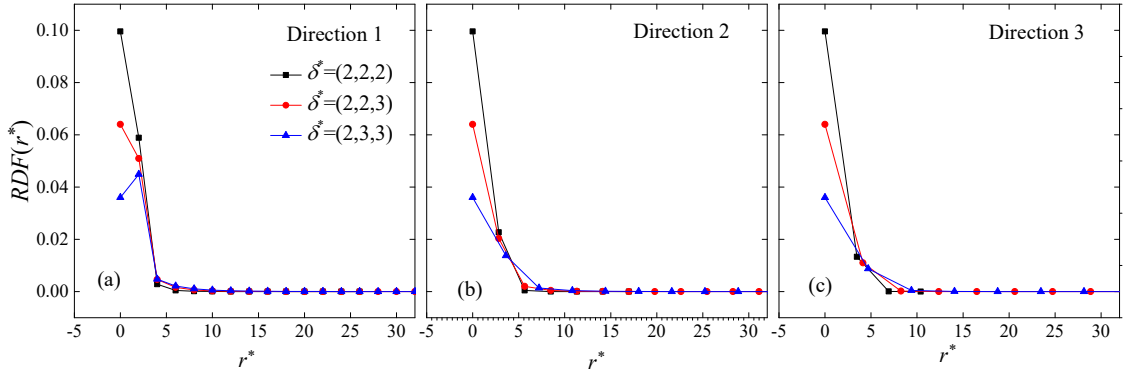

**Fig 18.** Variation of the RDF along three directions with particle wall emissivity of 0.4 when the cell edges are not equal: **(a)** direction 1, **(b)** direction 2, **(c)** direction 3.

## 4.3    Application of the continuum approach

Considering a densely packed particulate system shown in Fig. 19, the continuum approach proposed in this paper is applied to solve the temperature distribution in the system. Due to regular arrangement of particles, the applicability of the classic RTE for homogenization in this case is questionable. For simplicity, one chain of particles with periodic boundary (to consider the effect of surrounding particles) is considered to represent the particulate system, as shown in Fig. 19. The particles in the left and right boundary region are isothermal. The dimensionless particle spacing ($\delta^*$) is uniform. The temperature of particles located in the interior region needs to be determined. Totally 101 particles are placed in the interior region. The number of particles at the left or right boundary regions are set as 10, which is demonstrate to be sufficient. Hence the total number of particles ($N_P$) in the solution domain is 121.

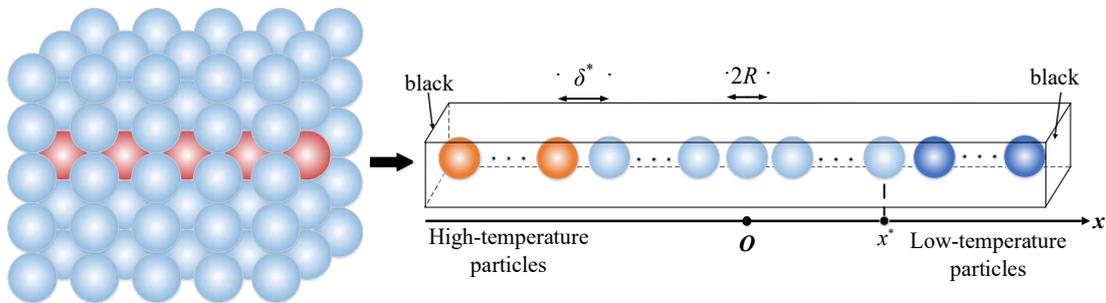

**Fig 19.** Schematic diagram of a non-random arrangement particulate system and one particle chain of the system is taken out to represent the entire particulate system.

The RDF is obtained by the interpolation with the singularity treatment with governing equation

Eq. (34). The temperature of high-temperature particles and low-temperature particles is fixed at 1000K and 300K, respectively. At first, a grid convergence test is performed with varying the number of subdivisions in the interior region ($N_{Int}$). Setting $\delta^*$=2.5 and $\varepsilon$ =1 for example, the RDF distribution and temperature along the linear chain are shown in <span style="color:red">Fig. 20 and 21</span>, respectively. The RDF is obtained by the interpolation from the RD with the singular point treatment presented in Section 2.2.3. Here, the cubic spline interpolation is used to obtain the RDF. The temperature distribution solved based on particle scale simulation is also shown for reference. It is shown that the temperature profile agree remarkably well with the results obtained by discrete scale simulation. The result can be considered converge when the $N_{Int}$ is greater than 61.

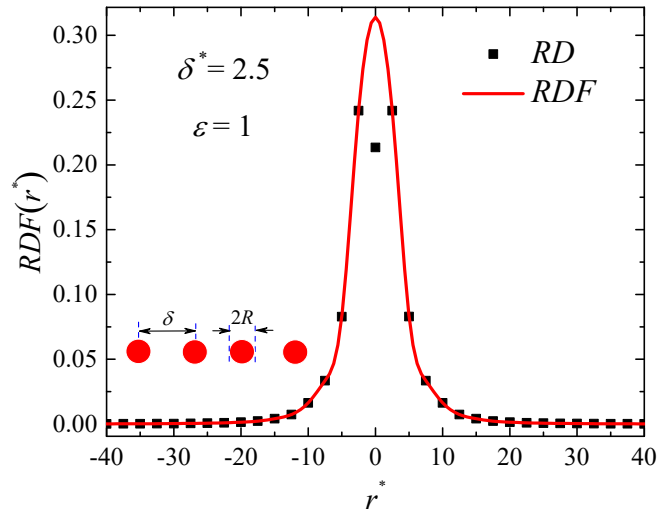

**Fig 20.** Fitted RDF distribution with singular point treatment.

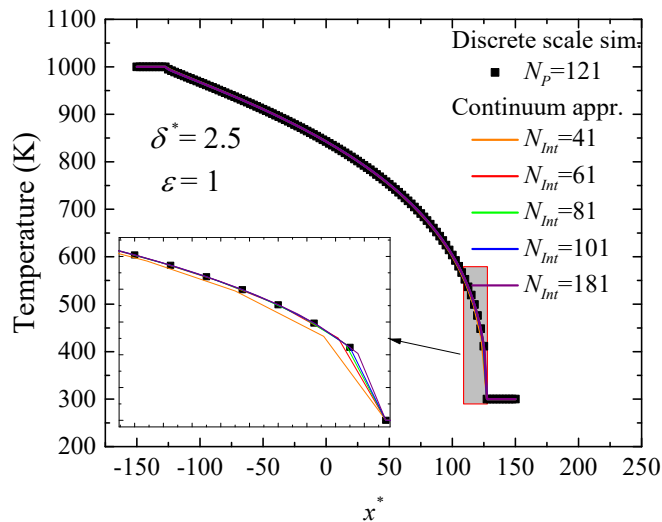

**Fig 21.** Grid convergence test for the continuum approach to obtain the temperature

distribution along the particle chain.

The RDF distribution and temperature along linear chain in the case of different particle spacing $\delta^*$ and particle wall emissivity $\varepsilon$ are shown respectively in Figs. 22 and 23. It is noted that the temperature distribution shows a strong nonlinear behavior. Near the high-temperature boundary, the temperature profile changes slowly, while a quick drop of temperature occurs near the low-temperature boundary. The reason is that the contribution of high-temperature particles on the temperature distribution is greater than the low-temperature particles due to nonlinear dependence of radiation power with temperature. As shown in Fig. 22, the decay rate of RDF increases when $\delta^*$ decreases. As a result, the temperature profile decreases more abruptly with $x^*$ for smaller value of $\delta^*$ decreases. As shown in Fig. 23, the decay rate of RDFs increases at larger $\varepsilon$, hence the temperature profile decreases more quickly at larger $\varepsilon$. It can be explained that small $\varepsilon$ will enhance multiple reflections among particles and reduce particle absorption at the same time, resulting in a spreading of the shape of RDFs.

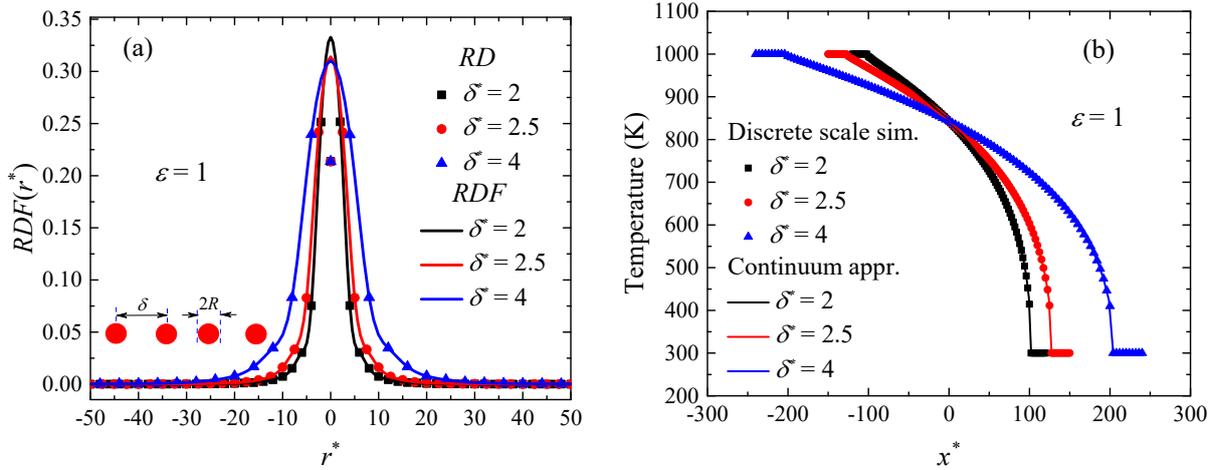

**Fig 22.** RDF and temperature distribution along linear chain for different particle spacing:

**(a)** RDFs, **(b)** temperature distribution.

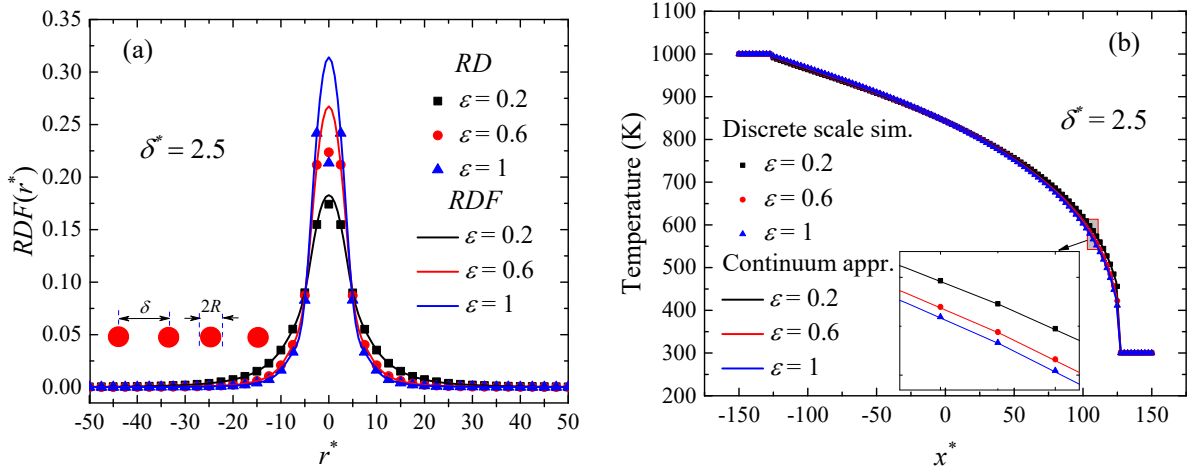

**Fig 23.** RDF and temperature distribution along linear chain for different particle wall emissivity:

**(a)** RDFs, **(b)** temperature distribution.

To illustrate the effect of the temperature difference ($\Delta T$) on temperature profile, three different value of $T_R$ are considered, namely, 300K, 500K and 700K, as shown in Fig. 24. The $\delta^*$ and $\varepsilon$ are fixed as 2.5 and 1, respectively. It is observed that the temperature distribution tends linear as $\Delta T$ decreases.

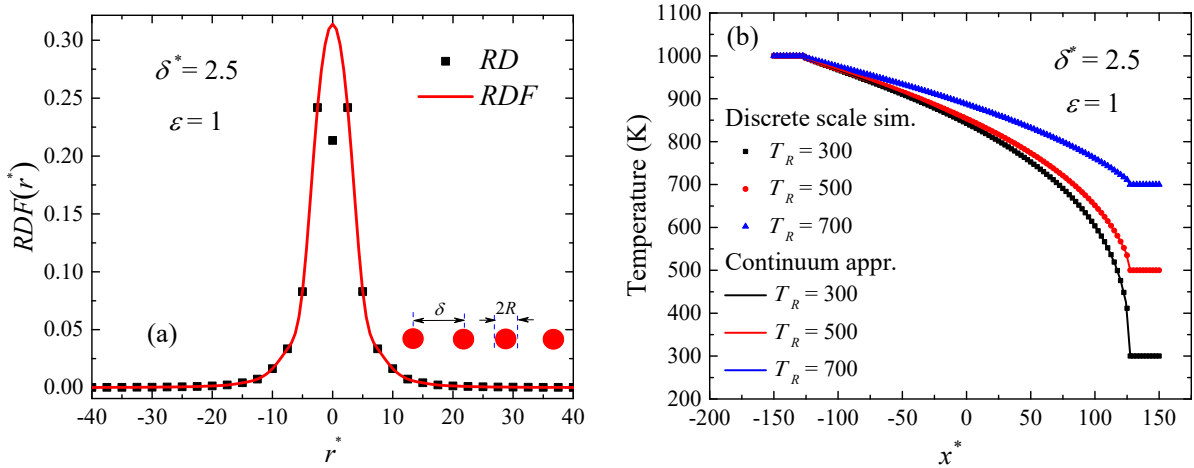

**Fig 24.** RDF and temperature distribution along linear chain for different $T_R$:

**(a)** RDFs, **(b)** temperature distribution.

Figure 25 shows the influence of the singular point treatment on the temperature distribution. As shown, when the number of subdivisions is large enough, i.e. $N_{\text{Int}}$ is greater than 61, the temperature distribution with the singular point treatment are the same as that without singular point treatment.

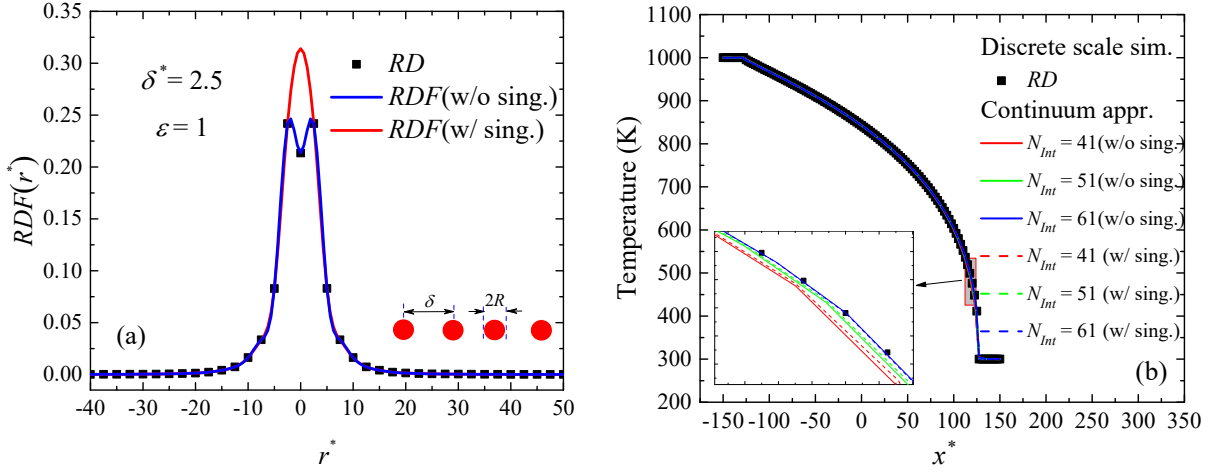

**Fig 25.** RDF obtained with the singular point treatment or not and the solved temperature distribution along the particle chain using the different RDFs: **(a)** RDFs, **(b)** temperature distribution.

Besides to obtain the RDF from interpolation, it can also be obtained by fitting from the RD. The effect of different methods of obtaining RDF on the solved temperature distribution is analyzed in the following. The case of $\delta^*$=2.5 and $\varepsilon$=1 is considered. Figure 26 shows the RDFs obtained through different methods and the corresponding solved temperature distribution using the different RDFs. The prediction model describing the trend of RDF is given as follows,

$$RDF(r) = (c_0 + c_1 r^1 + c_2 r^2)\frac{1}{a + r^2} e^{-\beta r} \tag{49}$$

in which, the fitting parameters $c_0$, $c_1$, $c_2$, $a$ and $\beta$ are determined as 2.46, -0.066, 0.0007, 1.47 and 0.039, respectively. The fitted RDF is presented in Fig. 26(a). As shown in Fig. 26(b), the temperature distribution solved using the fitted RDF are about the same as that solved using the interpolated RDF. Both results are in good agreement with the reference result solved through discrete scale simulation. It is noted that the interpolated RDF does not have a simple mathematic form, while the fitted RDF can be written by a simple formula, such as Eq. (49). Generally, the temperature distribution for all the cases solved based on the continuum approach are in remarkable agreement with the results obtained by discrete scale simulation. This demonstrates that the continuum approach has good performance to study radiation heat transfer in non-random particulate system.

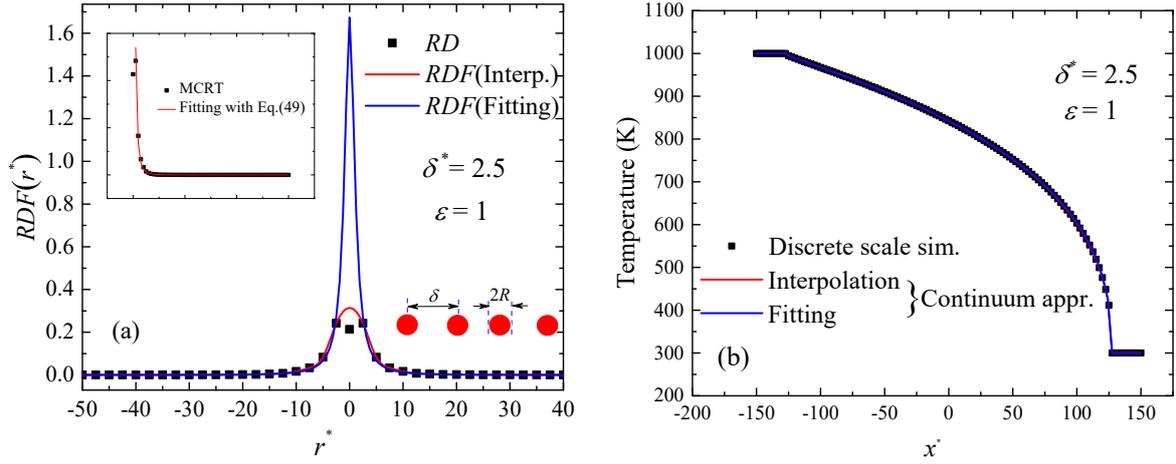

**Fig 26.** RDF obtained through different methods and the solved temperature distribution along the particle chain using the different RDFs: **(a)** RDFs, **(b)** temperature distribution.

## 5 Conclusion

The applicability of the RTE is questionable for densely and regularly packed particulate systems, due to dependent scattering and weak randomness of particle arrangement. A new continuum approach that does not explicitly rely on the RTE is proposed for radiative heat transfer in densely packed particulate system, which potentially overcomes the limitations of the RTE, namely, point scatterer assumption, and the assumption that the position of particles are completely uncorrelated. The obtained governing equation is in integral form, with RDF as the continuum scale physical parameter that characterizes the radiative transfer properties of the system. A theoretical model of the RDF is developed for the randomly distributed particulate system. The effect of surface scattering modes, filling ratio and particle wall emissivity of different particle arrangements (randomly or regularly) on the distribution of RDF is investigated. In a regularly distributed particulate system, the RDF distribution is shown to be anisotropic. Numerical experiments show that the temperature field solved based on the proposed continuum approach agrees remarkably well with that solved based on the particle-scale simulation, demonstrating good performance of the new continuum approach to study radiation heat transfer in non-random particulate system.

It is noted that the concept of RD is general, which can be applied to particles of any shape and with any kind of surface properties, such as diffuse or Fresnel reflection and can be transparent. In case that the self-emissive power of a particle is well defined, the RD can be defined. Hence the derivation presented in this work to obtain the continuum theory based on the RD is also general applicable.

# Conflict of Interest

The authors declared that there is no conflict of interest.

# Acknowledgement

The support of this work by the National Key Research and Development Program of China (No. 2018YFA0702301) and the National Natural Science Foundation of China (No. 51976045) are gratefully acknowledged.